# Interfacial structure in organic optoelectronics

A. Turak[1], D. Grozea[1], C.J. Huang[1], Z.H. Lu[1]

[1]Department of Materials Science and Engineering, University of Toronto, Canada

**Abstract**

The interfacial structure plays a critical role in modern optoelectronics. Currently multilayer electrodes are used to optimize the injection and lifetime properties. The choice of interlayer is not universal, with different effects for the same material with different capping metals. Using a novel *in-situ* characterization method with X-ray photoelectron spectroscopy, the organic/inorganic interface in OLEDs was examined for two common cathode metals with a LiF interlayer. The impact of the interfacial layer on the performance of devices can be attributed to the bulk lattice matching of the interfacial layer and of the by-products of interfacial oxidation, and the metallic cathode.

**Introduction**

Since the first stable multilayered organic light emitting diode (OLED) structure was produced [1], there has been widespread interest in such optoelectronic devices. Mainly due to potential applications in flat panel displays, many materials have been investigated in order to produce the entire visible spectrum [2]. These studies show that the interfacial region between the active layers and the electrodes plays a primary role in device performance. The introduction a thin, 5-10Å, ionic insulator, such as LiF, at the metal electrode/organic interface was found to substantially improve the device performance [3], in fact changing the conduction regime from space-charge limiting to injection limiting in some cases [4]. However, this improvement is not universal; some metal cathodes, such as Mg, show significant degradation of device properties with the introduction of a LiF interlayer [5]. While the effect of LiF is well documented, the underlying working principle is still poorly understood. Many of the proposed mechanisms do not adequately explain the effect observed for different cathode combinations. As well, most are insufficient to explain the operational capacity of devices with thick interlayers, which show no change in the injeciton properties compared with thinner interlayers. One possible explaination for these disperate effects can be related to the differing nature of interfacial oxidation for different cathodes. Recent work on the oxidation of LiF coated metal surfaces [6] indicates that while LiF slows down the oxidation of Al, it enhances and changes the oxidation by-products for Mg. In this investigation, we use X-ray photoelectron spectroscopy to explore the electrode/organic interface in some OLED devices with a LiF interlayer. The introduction of LiF does indeed supress interfacial oxidation for Al, and enhance interfacial molecular fragmentation for Mg.

**Experimental**

The OLED structures examined were produced using a Kurt J. Lesker OLED cluster tool. The configuration was based on multi-layer devices of glass/indium tin oxide (ITO)/ 600 Å N,N'-diphenyl-N,N'-bis(3-methylphenyl)1,1'-biphenyl-4,4' diamine (TPD)/250Å 8-tris (hydroxyl quinoline) aluminum ($Alq_3$). For the different cathode materials, additional layers were deposited atop this basic configuration. The first consisted of 150Å $Alq_3$/10Å LiF/ 2000Å Mg; the second of 200Å $C_{60}$/100Å LiF/1500Å Al. The samples were then prepared by an in-vacuum peel-off method, as described elsewhere [8]. This *in-situ* peel-off technique produces perfect cleavage at the metal/organic interface, converting the buried interface into an organic film surface and a cathode surface. The XPS spectra were generated by a Mg K$\alpha$ source with a photon energy of 1253.6 eV in an attached Phi ESCA 5500. For depth profiling analysis, the sputtering is performed using an 3 keV $Ar^+$ ion beam at 60° incidence angle.

**Results and Discussion**

**Mg/LiF devices**

Highly reactive Mg cathodes are known to cause molecular fragmentation reactions at the interface with $Alq_3$, the most widely used organic molecule [7]. Though it has been proposed that LiF could prevent such interfacial reactions [8], in fact such an interlayer appears to have enhanced such reactions. Figure 1 of the Al 2*p* core level for the cathode side of the interface clearly shows an asymmetric line shape consistent with interfacial breakdown [7]. The composition ratios also support the molecular fragmentation, with the Mg/LiF surface showing a significant N deficiency. As the theoretical N/Al ratio is 3 N per Al atom for $Alq_3$, N deficiency is an indication that the molecular structure is no longer consistent with $Alq_3$. On the cathode surface for both types of cathodes, the Mg 2p core level is highly asymmetric, corresponding to multiple oxidized and metallic states. Figure 2 shows the difference in the binding energy shift between the metallic and oxide peaks for the two surfaces, indicative of a different oxide phase forming at the interface with a LiF interlayer.

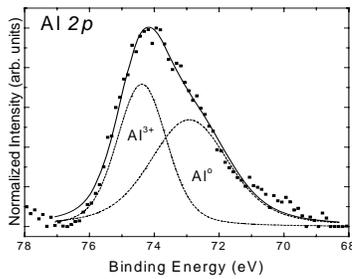

Figure 1 Al 2*p* core level recorded for the Mg/LiF surface
The experimental data can be well fitted by the sum (solid line) of two separate peaks (dashed lines), one metallic state at 72.7 eV and another $Al^{3+}$ state at 74.4 eV.

This is consistent with recent results on the oxidation of coated Mg surfaces, where LiF favoured the formation of $MgCO_3$ [6]. As the ion exchange reaction between Al and Mg causes molecular fragmentation, there would be an abundance of carbon based fragments at the interface, which would allow formation of such carbonates. Since both $MgCO_3$ and LiF have poor bulk lattice matching with Mg [6], the disruption of the interface with carbonate formation would account for the poor device performance.

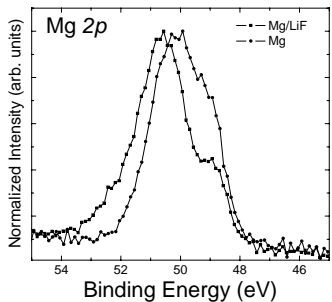

Figure 2 Mg 2*p* core level for both Mg and Mg/LiF cathodes.

**Al/LiF devices**

Unlike Mg/LiF, Al/LiF is an exceptionally useful cathode for OLEDs [3]. Investigation of interfacial oxidation of Al is difficult with $Alq_3$ as the oxidation state is the same for the cathode oxide and the molecule itself. $C_{60}$ is a very effective electron transport material when used with Al/LiF cathodes [9], and provides an ideal system to study interfacial oxidation as oxygen can only come from lateral diffusion. Though thin interlayers of LiF can radically improve conduction, such improvement has been observed even at interlayer thickness of 100Å [10]. This performance cannot be explained adequately by any currently accepted mechanism for interfacial injection enhancement. Figure 3 shows the Al visible at the cathode /organic interface after peel-off for both Al and Al/LIF cathodes. There is more visible oxidation, as shown by the high binding energy core level, for the Al cathode (figure 3(a)). However, the injection characteristics are controlled by the injection region, located at the metal surface. Figure 3 (c) again indicates greater oxidation, and three times the amount of oxygen, at the surface without a LiF interlayer. This leads to the observed injection limited conduction characteristics.

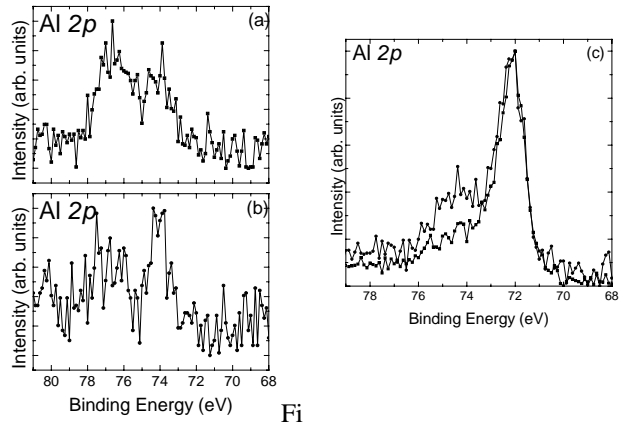

Figure 3 Al 2*p* core level for (a) Al surface (b) Al/LiF surface after peel-off (c) the injection region of the cathode after $Ar^+$ sputtering.

LiF, even in small quantities, provides an effective oxidation barrier on Al surfaces [6], due primarily to good bulk lattice matching. As the Al ions diffuse through the interlayer, they encounter diffusing oxygen atoms, and trap them away from the injection zone. This has the additional benefit of consuming some oxygen that can act as bulk conduction traps within the $C_{60}$ layer [4]. Therefore, the lattice matched interlayer encourages space-charge limited conduction by both scavenging oxygen within the LiF layer and preventing oxidation at the critical injection region.

**Conclusion**

The effectiveness of interlayers in multilayer cathodes can be related to the interaction between the interlayer and the cathode material. For Al, with good lattice matching, LiF provides interfacial oxidation resistance, explaining the space charge limited conduction observed for devices using such interlayers. For Mg, where the interlayer not only has poor lattice matching but also encourages formation of poorly matched oxidation by-products, the enhanced interfacial oxidation and molecular fragmentation would explain the reduction in the device properties for such devices.